\pdfoutput=1

\documentclass[11pt]{article}

\usepackage{acl}
\usepackage{times}
\usepackage{latexsym}

\usepackage[T1]{fontenc}

\usepackage[utf8]{inputenc}

\usepackage{microtype}

\usepackage{inconsolata}

\usepackage{amsmath,graphicx}
\usepackage{amssymb}
\usepackage{multirow} 
\usepackage{url}

\usepackage{hyperref}

\title{MusiLingo: Bridging Music and Text with Pre-trained Language Models for Music Captioning and Query Response}

\author{Zihao Deng $^{1}$\thanks{$^*$The authors contributed equally to this work. 
    Emmanouil Benetos is corresponding author.
    }\quad  Yinghao Ma\textsuperscript{2,\includegraphics[scale=0.03]{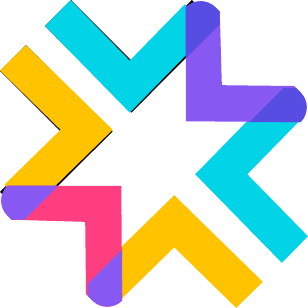}*}\quad Yudong Liu$^{3}$\quad  Rongchen Guo$^{4}$
    \quad Ge Zhang$^{\includegraphics[scale=0.03]{images/map-logo-c.pdf}, 5,6}$ \\
    {\bf Wenhu Chen$^5$}\quad  {\bf Wenhao Huang$^{6}$}\quad {\bf Emmanouil Benetos$^{2\dag}$}\\
    $^1$ University of Pennsylvania, USA \\
    $^2$ Centre for Digital Music, Queen Mary University of London, UK\\
    \includegraphics[scale=0.03]{images/map-logo-c.pdf} Multimodal Art Projection Research Community\\
    $^3$ Carnegie Mellon University, USA\quad     $^4$ University of Texas at Austin, USA\\
    $^5$ University of Waterloo, Canada. \quad    $^6$ 01.AI, China\\
 }

\begin{document}
	\maketitle

\begin{abstract}
Large Language Models (LLMs) have shown immense potential in multimodal applications, yet the convergence of textual and musical domains remains not well-explored. To address this gap, we present MusiLingo, a novel system for music caption generation and music-related query responses. MusiLingo employs a single projection layer to align music representations from the pre-trained frozen music audio model MERT~\cite{li2023mert} with a frozen LLM, bridging the gap between music audio and textual contexts. We train it on an extensive music caption dataset and fine-tune it with instructional data. Due to the scarcity of high-quality music Q\&A datasets, we created the MusicInstruct (MI) dataset from captions in the MusicCaps datasets, tailored for open-ended music inquiries. Empirical evaluations demonstrate its competitive performance in generating music captions and composing music-related Q\&A pairs. Our introduced dataset enables notable advancements beyond previous ones.
	\end{abstract}

\section{Introduction}
In the realm of Music Information Retrieval (MIR), prevailing methodologies for contemporary musical descriptions typically lean on discriminative learning. An illustrative instance is music tagging \cite{law2009evaluation, won2020evaluation,  won2021semi}, where descriptors encompassing genres, composers, instruments, emotions, and tempos are ascribed to each music clip. 
In this case, the model output is confined to a pre-determined set of categorical labels, thereby constraining its applicability in contexts like music exploration and recommendation, where the ability to handle and generate handle natural language descriptions instead of individual tags such as music captions or answers to music instructions would boast a diverse array of practical applications. 
These include generating textual descriptions for items found within extensive music catalogues, annotating copious user-generated content; automatically providing descriptions for evocative music featured in videos, catering to the needs of the hearing-impaired; and furnishing explanations for automated music recommendations. 
Furthermore, this advancement facilitates enhanced search and discovery of musical material for composers, all through user-friendly queries, while also serving as an inspiration for text-based music generation algorithms.

Given the potential alignment between musical and textual representations, there is some research to bridge the gap between acoustic music and natural language modalities, though still in its relatively nascent stages. 
A prospective avenue for better performance, in light of the recent triumphs of large language models (LLMs), entails integrating the conversation and generalisation proficiencies offered by LLMs into musical tasks.

Considering these insights, we introduce a novel music language model designed for music captioning, question answering, and query responses. Our approach involves a single projection layer configuration with temporal compression applied to music embeddings. 
In contrast to the multilayer perceptron (MLP) approach for the Llama-adapter~\cite{zhang2023llama} in a contemporary work MU-LLaMA~\cite{liu2023music}, which projects music embeddings to the upper layers of Llama, our method employs a straightforward projection to convey the embeddings to the initial layer of Llama. 
The Llama adapter is designed for finetuning the vanilla language model Llama into an instruction-following model and enhances its applicability to visual-language tasks but Vicuna~\cite{vicuna2023}, the LLM we use as a language decoder, is already capable of instruction-following.
Our simple approach offers the advantage of accommodating larger batch sizes which possess an NLP backbone capable of instruction-following tasks. This has demonstrated considerable efficacy in visual-language contexts, as evidenced by its successful implementation in models like Llava~\cite{liu2023visual} and mini-GPT4~\cite{zhu2023minigpt}. 

We also incorporate a pre-training phase to align music information with textual representations, utilising a large amount of music captioning data, and fine-tuning the model using our developed MusicInstruct dataset derived from GPT-4~\cite{brown2020language}. 
This equips our model with the capability to understand different aspects of musical compositions and enables it to provide accurate and natural responses to user queries.

In summary, our work features the following core contributions:
\begin{itemize}
\item We introduce MusiLingo, a novel music-language model capable of performing music question answering and captioning; 
\item We demonstrate superior performance and state-of-the-art (SOTA) modelling for a variety of metrics for music Q\&A; 
\item We create a new MusicInstruct (MI) dataset, which features 60,493 Q\&A pairs covering both general questions like music summarisation, and specific questions related to music genres, moods, and instruments.
\item Our ablation study delves into the impact of fine-tuning datasets on MusiLingo's performance. It reveals that the choice of training data significantly influences the model's effectiveness. 
\end{itemize}

Section 2 details our methodology for the MI dataset creation and the music question-answering tasks. Section 3 outlines the MusiLingo model structure and training procedure.  Section 4 presents experiments and evaluations of our model and baselines. Our code is available on GitHub\footnote{\href{https://github.com/zihaod/MusiLingo}{GitHub Repository}}.

\section{Related Work}
Several prior studies have explored the alignment of acoustic audio modalities with NLP. ~\citet{cai2020music} employed a CNN encoder based on a spectrum alongside an RNN with an attention-based natural language decoder to predict tag lists. However, the evaluation metrics utilized were not comprehensive, and the resulting tag lists exhibited noise, diminishing the robustness of the study. 
~\citet{doh2023universal} utilized CLove or BERT along with transformer architectures for music encoding to facilitate text-to-music retrieval. MusCALL \cite{manco2022contrastive} leverages contrastive learning for text-to-audio and audio-to-text retrieval tasks. \citet{wu2023largescale} curated a large-scale audio caption dataset comprising 630k samples and trained models utilizing various pre-trained audio encoders and natural language decoders, also equipped with contrastive learning capabilities. Their approach yielded competitive results in text-to-audio retrieval tasks.

In addition to raw audio processing, ~\citet{doh2021music} utilized transformer and GRU models to map song IDs and metadata from each playlist to the corresponding playlist titles. \citet{kim2023music} incorporated artist IDs as part of the input for their investigation.

Work has additionally been carried out on music and audio captioning. MusCaps~\cite{manco2021muscaps} leverages convolutional networks for music understanding and recurrent neural networks for captioning. 
MuLan~\cite{huang2022mulan} uses contrastive learning to align the text embedding to audio representations for music tagging and retrieval of music with text query. But the work is not open-sourced.
LP-MusicCaps~\cite{doh2023lp} and audio captioning transformer (ACT)~\cite{mei2021audio} utilise a cross-modality transformer-based encoder-decoder architecture for music/audio captioning. \citet{choi2017music} utilized an RNN to map CNN embeddings of playlists to word2vec embeddings of playlist captions. However, the performance of models in this earlier study is far from perfect. Additionally, PLAYNTELL~\cite{gabbolini2022dataefficient} employs tags, artist distributions, and audio as input for playlist captioning, thereby achieving state-of-the-art performance.
Although these studies have shown notable advancements in tackling music captioning, they are not designed for music instruction-following and their effectiveness in functioning within a genuine conversational context for question-answering remains restricted or not evaluated. 

Several works have applied LLMs to multimodal tasks. UniVAL~\cite{shukor2023unified} offers a versatile model for image, video, audio, and language modalities, while LTU~\cite{gong2023listen} excels in audio quizzing. However, none of these models are designed for instruction-following on general audio, making them unsuitable for music-related question-answering and dialogue especially polyphonic music-related topics such as key and chord.
To enable the bridge of two modalities on limited resources, we are inspired by the success of vision-language pre-training. In vision-language pre-training, the prevailing approach is to follow a new paradigm, connecting pre-trained unimodal encoders with LLMs via a learnable interface. This approach keeps encoders and language models fixed, using query tokens or adapter layers~\cite{zhang2023llama} to transfer information between modalities.
The interface can be a set of query tokens that extract information from the modality, as BLIP-2~\cite{li2023blip} and Flamingo~\cite{alayrac2022flamingo}, or an adapter layer that projects embeddings from one modality to another. Mini-GPT4~\cite{zhu2023minigpt} and Video-ChatGPT~\cite{maaz2023videochatgpt} use simple linear adapters to project the visual embeddings onto text embedding space. Video-LLaMA~\cite{zhang2023videollama} adopts the Q-Former design from BLIP-2 for the adapter and incorporates 2 projections from the image and audio data in the video. 
LLaMA-Adapter~\cite{zhang2023llama} employs a parameter-efficient approach with small adapter modules within transformer blocks. 
A contemporary work, MU-LLaMA~\cite{liu2023music}, extends the LLaMA-adapter concept to music language tasks. These models, which utilise pre-trained frozen encoders and learnable interfaces, offer a promising approach to connecting any modality with language models, providing efficient training and maximal preservation of the model's original knowledge.

\section{Dataset \& Evaluation Metrics}\label{Dataset}
\subsection{Large Dataset for Pre-training}
In our study, we utilise the LP-MusicCaps-MSD dataset \cite{doh2023lp} for pre-training. This dataset is derived from the ECALS subset \cite{doh2023toward} of the Million Song Dataset \cite{bertin2011million} and consists of 520k 30-second clips with a vocabulary of 1054 labels encompassing various categories such as genre, style, instrument, vocal, mood, theme, and culture. Each music clip is associated with an average of 10.2 labels, used for generating pseudo captions, including one caption, one summary, and one rephrased version for each audio clip using the GPT-3.5 model. We employ this extensive GPT-generated dataset for pre-training and subsequently fine-tune our results using a smaller, high-quality Q\&A dataset.

\begin{figure*}[htb]
\centering
\includegraphics[width=\linewidth]{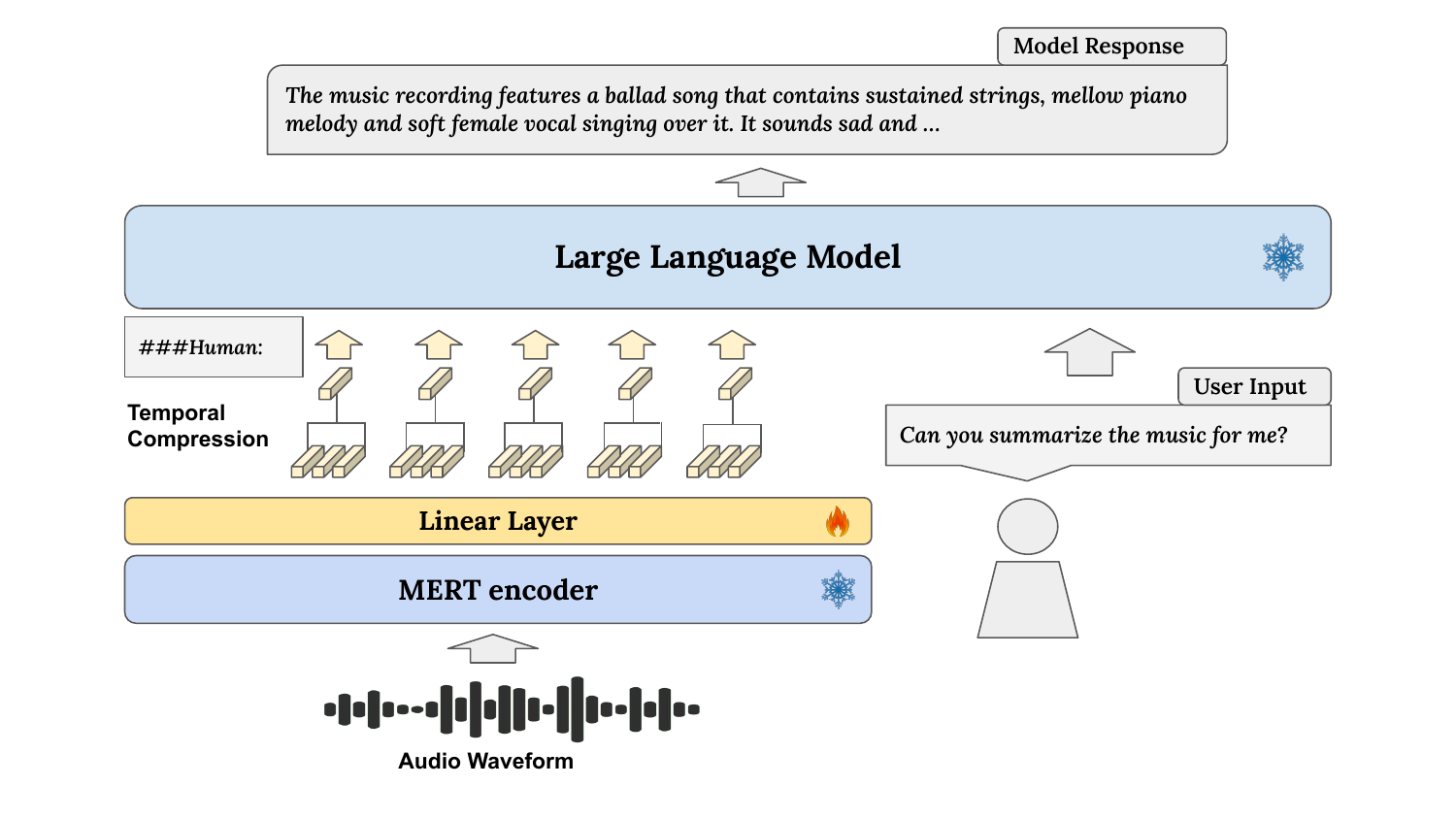}
\caption{Overview of the MusiLingo model. Note that the backbone LLM can be easily replaced from Vicuna-7B to other LLMs.}
\label{fig:model}
\end{figure*}

\subsection{Music Instruction Following Dataset}
\subsubsection{Collection Process}

To enhance the model's ability to generate content of superior quality, we conducted additional fine-tuning using a bespoke music Question-Answering dataset we developed and named the MusicInstruct (MI) dataset. This dataset comprises Q\&A pairs corresponding to individual musical compositions and is expressly tailored to tackle open-ended inquiries within the realm of music. It is derived from the music-caption pairs in the MusicCaps dataset~\cite{agostinelli2023musiclm}.
The dataset is released with \texttt{cc-by-nc-4.0} license. The audio is available on YouTube with the given id, and the Q\&A pairs along with metadata can be downloaded at our Huggingface page \footnote{{Download dataset at } \url{https://huggingface.co/datasets/m-a-p/Music-Instruct/tree/main}}

The MI dataset was constructed through prompt engineering and the application of few-shot learning techniques to ChatGPT~\cite{openai2023gpt4}.
Given the ground truth caption of a musical excerpt from the MusicCaps dataset, we design a prompt instructing the chatbot to generate multiple Q\&A pairs based on the provided caption. 
An example prompt for ChatGPT generating Q\&A pairs is given in Table ~\ref{eg}.
The prompt consists of three parts:
(1) An instruction delineating the task, serving as a system message directed at ChatGPT;
(2) A set of few-shot example questions that the chatbot may generate; and
(3) A concluding query featuring the music caption in question.

After generating all Q\&A pairs, we employ another prompt to categorize whether the generated Q\&A pair accurately encapsulates the essence of the music caption (e.g., Does this question-answer pair come from the context delimited with \#\#\#\#?). Pairs that ChatGPT classified as negative were filtered out.
In addition, some other problematic Q\&A pairs have also been removed from the MI dataset,
including generations with runtime errors and instances where the generation terminates improperly (i.e., lacking punctuations in the end).

The resulting MI dataset comprises two versions, spanning questions from the general (\textit{v2}) to the specific (\textit{v1}).
\textit{v1} encompasses 27,540 Q\&A pairs, with detailed questions and one or two-sentence long concise answers.
The questions delve into various detailed aspects, such as music tempo, mood, instruments used, singer, genre, and music tags - important attributes and properties of a piece of music. 
Conversely, \textit{v2} encompasses 32,953 Q\&A pairs, featuring general questions, with answers typically being more extensive and serving as paraphrased renditions of the original caption.
\textit{v2} reflects a broader overview of the music content.
Appendix A lists some example Q\&A pairs from \textit{v1} and \textit{v2} of the MI dataset.

\subsubsection{Quality Evaluation}
\begin{table}[htb]
\centering
\resizebox{.75\width}{!}{
 \renewcommand{\arraystretch}{1.3}
\begin{tabular}{l|lll}
\hline\hline
 Model & MusicQA & MI short & MI long\\
\hline
Instruction has Clarity & 83.3\%	&93.0\%&100.0\%\\
Instruction has Feasibility	& 81.5\%	& 95.6\%&	99.7\%\\
Instruction has Practicality & 83.3\%	&93.3\%&	100.0\%\\
Output Quality excellent & 70.4\%	&95.6\%& 97.5\%\\
Output Quality not failed & 94.4\%	&99.2\%& 100.0\%\\

\hline\hline
\end{tabular}
}
 \caption{Quality Assessment of Instruction Clarity, Feasibility, Practicality, and Output Quality of MusicQA and MusicInstruct dataset.}
  \label{tab:quality}
\end{table}

We conducted a comprehensive quality assessment of the MusicInstruct dataset Q\&A pairs utilizing the assessment method in Kun~\cite{zheng2024kun}. For both short and long versions of the dataset, we randomly selected 1\% of the instruction pairs, thus evaluating 600 pairs. We annotate the dataset quality on our own on whether the instruction fulfils the following three properties: Clarity, Feasibility and Practicality, therefore revealing the overall quality of the instruction.
Additionally, we used the ``consistency'' of the responses and the related instruction to examine the quality of the outputs. Evaluators were asked to rate each output as excellent, passed, or failed, depending on the extent to which it met the requirements and intentions of the institution. See \ref{criteria} for definitions of the evaluation metrics.

\subsection{Evaluation Metrics}
Both music captioning and music question answering are text-generation tasks.
To this end, we use well-established text generation metrics to evaluate the model performances on both tasks, 
where the generated music captions/Q\&A are compared to the ground truth texts.
Metrics we used include BLEU~\cite{Papineni02bleu:a, lin-och-2004-orange}, METEOR~\cite{banerjee2005meteor}, ROUGE~\cite{lin-2004-rouge}, and Bert-Score~\cite{bert-score}. 

To make our results comparable with MU-LLaMA, we use the average of $BU_1$, $BU_2$, $BU_3$, and $BU_4$ as the result of the BLEU value.

\section{Method}
In this section, we introduce MusiLingo, a potent music-language model that leverages LLM capabilities to enhance music comprehension. The model's key innovation lies in the use of simple adapters, a prevalent technique in LLM-based multimodal models. Our approach builds upon a design where both the music encoder and LLM remain fixed, while a single adapter network is trained to project music embeddings into the text embedding space. As demonstrated in fig. \ref{fig:model}, We utilised MERT-330M \cite{li2023mert} as the music encoder and Vicuna-7B \cite{vicuna2023} as the language model, with the adapter consisting of a simple linear layer followed by temporal compression. Our methodology involves pre-training and instruction tuning to grasp music concepts and generate coherent responses. This streamlined design substantially reduces the time and resources needed for music-language model training.

\subsection{Model Architecture}
There have been a variety of designs for how and where to use adapters \cite{zhu2023minigpt, li2023blip, alayrac2022flamingo, maaz2023videochatgpt, zhang2023videollama, zhang2023llama, liu2023visual}, and our work extends from the design where both the music encoder and the LLM are completely frozen, and one adapter network is trained to project music embeddings onto the text embedding space. This adapter design, which has demonstrated remarkable efficacy in vision-language models like Llava \cite{liu2023visual} and Mini-GPT4 \cite{zhu2023minigpt}, enables us to easily insert music features into text embeddings and wrap it around with the user questions and necessary prompts that are used by different language models for doing instruction-following tasks. It also allows us to use larger batches during training. We used MERT-v1-330M \cite{li2023mert} as our music encoder and Vicuna-7B \cite{vicuna2023} as the language model. MERT \cite{li2023mert} is a self-supervised music understanding model which employs teacher models to generate pseudo labels for sequential audio clips during training. It features a multi-task paradigm that simultaneously learns the musical and acoustic representations of the input music, thus achieving SOTA performances on various music information retrieval tasks. Vicuna \cite{vicuna2023} is a chat model fine-tuned upon LLaMA \cite{touvron2023llama} using 70K user-shared conversations, showing better performance than open-source language models like LLaMA \cite{touvron2023llama}. Our adapter is a simple linear layer followed by a temporal compression operation. We perform both a pre-training step and an instruction tuning step to learn the music concepts and form them into coherent answers. This simple yet effective design significantly reduces the time and resources needed to train a music-language model and helps bridge the gap between these two modalities.

The MusiLingo model consists of a music encoder, an adaptation layer, and a pre-trained LLM to achieve cross-modal understanding between music and text data. In particular, 
We use MERT as our music encoder to extract the acoustic and musical information from the input music clip and use Vicuna as the language model, which takes the music embedding output from the adaptation layer and generates text responses based on additional user text input. For the adaptation network we use a simple linear layer, which has been demonstrated to be fairly effective in a few recent works in the vision-language domain \cite{ maaz2023videochatgpt, liu2023visual, zhu2023minigpt}. Note that the choice of a linear layer is also based on the observation that MERT has encapsulated the information in different dimensions via its attention layers. Consequently, there may not be an imperative need to introduce supplementary architectural elements, such as attention layers or BLIP-2 Q-Former \cite{li2023blip}, for the acquisition of temporal dimension information.

To harness both high-level and low-level information within music audio, we calculate the final music embedding by taking the weighted average of the outputs from each transformer block in the MERT model. This embedding is then projected onto the text embedding space of the language model via a linear layer.
However, the encoded music representations can be lengthy, posing training challenges, and the uncompressed sequence elements lack meaningful alignment with the language model's token embeddings. To address this, we introduce a temporal compression step following the linear layer. Given the output embedding $M \in \mathbb{R}^{B \times T \times D}$ from the adaptation layer (with $B$, $T$, and $D$ representing batch size, number of timesteps, and embedding dimension, respectively), we compress subsequences of length $t$ along the temporal dimension by computing the average. This results in a new embedding with a reduced temporal dimension of $T' = \lceil T/t \rceil$. Thus, the input to the language model after compression is a vector of shape $B \times T' \times D$.

\subsection{Music-Text pre-training}
To train the MusiLingo model, we initiate a pretext task focused on aligning music concepts with the language model. In this phase, our goal is to effectively transform music embeddings into text embeddings using established music captioning datasets, specifically LP-MusicCaps-MSD \cite{doh2023lp}. As illustrated in Fig.\ref{fig:model}, each music clip undergoes encoding by the MERT encoder and the adapter layer for each music-caption pair. The ground truth caption is tokenised and converted into text embeddings using the Vicuna model, then appended to the music embeddings via concatenation. The loss is the original language modelling loss from the Vicuna model, with the tokens for regression limited to the caption tokens. This pre-training step is crucial for enabling the model to comprehend music concepts and convert them into textual representations.

\subsection{Music Instruction Tuning}
While the pre-training step plays a pivotal role in aligning music and text concepts, it alone does not suffice for generating high-quality conversational content. Hence, we incorporate an instruction tuning step to facilitate the model's ability to respond to various music-related questions. This fine-tuning process draws from two datasets: MI (detailed in Section \ref{Dataset}), and MusicQA \cite{liu2023music}, which contains question-answer pairs generated with the assistance of an LLM. Instruction tuning on these two datasets effectively imparts the model with the capability to answer music-related questions in a human-like manner and equips it with the knowledge to generalise to unseen tasks concerning musical content.

\section{Experiment and Results}
In this section, we introduce the experimental setup as well as present an evaluation of our model's performance on the Question-Answering of music on the MusicQA and MI datasets. Besides, we evaluate the performance of music captioning on the MusicCaps dataset. We compare our results to state-of-the-art models and discuss the unique challenges posed by this dataset. Last, we carry out an ablation study on training on different parts of the MI dataset.
\subsection{Experiment Setup}
In the experiments we compare our model against three other music-language models including LTU \cite{gong2023listen}, LTU-AS \cite{gong2023joint}, and MU-LLaMA \cite{liu2023music}. LTU \cite{gong2023listen} is a general audio-language model based on audio encoder, LLM, and LoRA \cite{hu2021lora}. LTU-AS \cite{gong2023joint} improves upon LTU by integrating the Whisper model for obtaining spoken text and enabling more general audio understanding. MU-LLaMA \cite{liu2023music} is another baseline which uses the same MERT \cite{li2023mert} encoder as ours, but with a design similar to LLaMA-Adapter \cite{zhang2023llama} for aligning music and text information. 

For our model, during the pre-training phase, we train the network by concatenating the encoded caption with the projected music embedding and optimizing it for the caption tokens using the original language modelling loss. To ensure consistency, we use only the "caption\_writing" in the pre-training dataset as the ground truth music caption since it contains mostly rephrased versions of each other.
For instruction tuning, each data instance consists of an instruction or music-related question and its corresponding answer. We concatenate the instruction text token embeddings with the music embeddings, and the answer token embeddings with the instruction embeddings, with an additional prefix \textit{\#\#\#Assistant:} denoting the start of the answer. The objective is language modelling, with only the answer tokens contributing to the loss computation.
During pretraining, we trained the model with a batch size of 32 for 20k steps using 4 A100 80G GPUs for 1-2 days. For each fine-tuning stage on different datasets, we completed 2 epochs of training on a single A100 40G GPU for 0.5-1 day. Please refer to our Github repo for detailed information on hyperparameters.

\subsection{Result Analysis on Question-Answering}

\begin{table}[htb]
 \begin{center}
 \resizebox{.6\width}{!}{
 \renewcommand{\arraystretch}{1.3}
 \begin{tabular}{l|llll}
 \hline\hline
 Model & B-U↑ & M-R↑ & R-L↑ & BERT-S↑ \\
 \hline\hline
 & \multicolumn{4}{c}{MusicInstruct (Short)}\\
\hline
LTU~\cite{gong2023listen} &  29.7  & 36.6  & 42.8 &  90.3\\ 
LTU-AS~\cite{gong2023joint} &  30.4  & 36.3  & 42.0 & 90.9\\ 
MU-LLaMA \cite{liu2023music}  & 45.5$^*$& 50.1$^*$& 51.3$^*$& 93.2$^*$ \\
MusiLingo / MI(short) & 47.0& 51.4&51.4 & \textbf{92.9}\\
MusiLingo / MusicQA + MI(short) & \textbf{47.1}& \textbf{51.7}&\textbf{51.6}& \textbf{92.9}\\

\hline\hline

 & \multicolumn{4}{c}{MusicInstruct (Long)}\\
\hline
LTU~\cite{gong2023listen} &  6.7  & 9.3  & 9.0 &83.1\\ 
LTU-AS~\cite{gong2023joint} & 6.0  &8.8  &8.2 &83.3\\ 
MU-LLaMA \cite{liu2023music} & 14.3$^*$& 25,6$^*$& 41.1$^*$& 88.6$^*$\\  
MusiLingo / MI(long) & \textbf{45.0} & \textbf{25.0}& \textbf{22.9}& \textbf{86.1}\\
\hline\hline

 & \multicolumn{4}{c}{MusicQA}\\ \hline
LTU~\cite{gong2023listen} &  24.2  & 27.4  & 32.6 &  88.7\\ 
Llama-adapter~\cite{zhang2023llama} &27.3 &33.4 &41.3 &89.5\\
MU-LLaMA \cite{liu2023music}  & 30.6  & \textbf{38.5}  & \textbf{46.6}  & 90.1\\ 
MusiLingo / MusicQA & 32.4 & 37.2 & 45.3 & 90.6\\
MusiLingo / MI short + MusicQA& \textbf{33.2} & 38.4& 46.5&  \textbf{91.0}\\ % 507978, 508374

\hline\hline
 \end{tabular}
}
\end{center}
 \caption{Music question answering results on the MI datasets and MusicQA. If the audio corresponding to the evaluation dataset is present in the pre-training dataset of a model along with caption information, the performance of instruction-following during evaluation may be overestimated. In such cases, we denote this potential overestimation by marking the corresponding entries in the table with a $\star$. }
 \label{tab:compare}
\end{table}

Table \ref{tab:compare} demonstrates the experimental results of various models in the field of music question answering. These are categorised into three different scenarios: ``MusicInstruct (Short)'' which represents the short questions on MI datasets, ``MusicInstruct (Long)'' which refers to the long subjective questions on the MI dataset, and ``MusicQA'' which denotes the test set of the MusicQA dataset generated from the tags of MTG-jamendo datasets\cite{bogdanov2019mtg}. The table presents performance metrics for four key evaluation criteria: B-U (Bleu-Uni), M-R (METEOR-Rouge), R-L (ROUGE-L), and BERT-S (BERT-Score). 

From the table, MusiLingo demonstrates the highest overall performance on MusicQA datasets.
``MusiLingo / MusicQA'' represent the model fine-tuned with Q\&A pairs on the finetune set) of the MusicQA dataset, generated from the MagnaTagaTune (MTT) dataset \cite{law2009evaluation}. Our experiments on the MusicQA dataset demonstrate competitive performance, aligning with the state-of-the-art (SOTA) results provided by MU-LLaMA. Specifically, our model achieves comparable performance on M-R and R-L metrics and surpasses the SOTA methods on BU and BERT-S, confirming its effectiveness in addressing the challenges posed by the Music question-answering task. 
Besides, ``MusiLingo / MI Short + MusicQA'' is finetuned on the short-question partition on the MI dataset and then is finetuned on the MusicQA dataset. The results are particularly excellent in the B-U and BERT-S metrics and have no significant difference in M-R and R-L compared to the SOTA approach. 

Furthermore, MusiLingo demonstrates more competitive results on MI datasets in terms of both short objective questions and long subjective questions.
In the objective question scenario, we see that ``MusiLingo / MI (Short)'' has achieved the highest scores for all rule-based evaluation criteria, outperforming other audio Q\&A models, and provides competitive results compared to MU-LLaMA. Moreover, ``The MusiLingo / MusicQA + MI (Short)'', doing the continuous training on ``The MusiLingo / MusicQA'', only demonstrates slight improvement.

In the long-form music instructions, ``MusiLingo / MI (Long)'' outperforms other models by a significant margin.
It is interesting to note that audio Q\&A baseline systems LTU~\cite{gong2023listen} and LTU-AS~\cite{gong2023joint} perform well on objective questions such as instrument events and genres, while performing poorly in this scenario, suggesting the effectiveness of the MusiLingo approach for handling queries with more extended and higher-level music semantics. 
Note that MU-LLaMA may not be a good baseline system for the query-response on the MI dataset due to label leak issues. The MU-LLaMA is trained on the pre-training partition of MusicQA, which includes audio recordings in the evaluation split of MusicCaps along with the MPT-7B-generated Q\&A pairs based on these recordings. The testing split of the MI dataset is based on the same audio in the evaluation split of MusicCaps along with the GPT-4-generated Q\&A pairs based on these recordings. Both Q\&A pairs include information on instruments, genre, emotion, singers, and the audience's feelings.

Overall, the experimental results suggest that MusiLingo is a promising model for music question answering, showing competitive performance across various scenarios. It is particularly strong in handling complex, long-form queries, making it a valuable tool for music enthusiasts and professionals looking for detailed and accurate answers to their questions. 

\subsection{Result Analysis on Music Captioning}

We investigate the effectiveness of utilising a pipeline approach for music captioning, shedding light on its potential benefits. Given some previous Q\&A models, such as MU-LLaMA which can perform captioning, we use the question ``Please give a caption to the music'' and the caption ground truth to train a music captioning model.
Our experiments are conducted on the MusicCaps dataset, and we present key performance metrics in Table \ref{tab:caption}.

We did not include MU-LLaMA in the table because MU-LLaMA uses the whole MusicCaps dataset audio for training and then evaluates the results on the private dataset, making comparisons with such models on the MusicCaps dataset as a testing set not entirely suitable. Besides, it lacks transparency in explaining its captioning process, with the opacity stemming from the inherent diversity in the prompts query.
\begin{table}[htb]
\centering
\resizebox{.65\width}{!}{
 \renewcommand{\arraystretch}{1.3}
\begin{tabular}{l|llll}
\hline\hline
 Model & B-U↑ & M-R↑ & R-L↑ & BERT-S↑ \\
\hline
MusCaps  \cite{manco2021muscaps}  & 10.2& 17.0 &22.2 & 83.5 \\
% BART \cite{lewis2020bart} %MuLaB
% & 13.8 & 20.7& 19.3& 87.1 \\
LTU~\cite{gong2023listen} &  4.6  & 7.6  & 8.5 &83.6\\ 
LTU-AS~\cite{gong2023joint} & 4.0  &6.0  &6.3 &82.9\\ 
LP-MusicCaps~\cite{doh2023lp}& 14.7 & \textbf{22.4} & 21.5& \textbf{87.8} \\
MusiLingo Pre-trained & 4.7& 6.5& 6.7& 80.7\\
MusiLingo / MusicCaps & \textbf{30.8} & 21.6& \textbf{21.7}& 86.8\\
\hline\hline
\end{tabular}
}
 \caption{Music captioning results on the MusicCaps datasets.}
  \label{tab:caption}
\end{table}

Table \ref{tab:caption} summarises the results obtained by various models on the MusicCaps dataset. 
These results underscore the effectiveness of our proposed Q\&A pipeline approach in improving music captioning performance. MusiLingo provides SOTA performance in B-U and R-L metrics. 
However, we acknowledge that our model's performance in music captioning is still not on par with the current SOTA models, especially on the BERT-score. Further improvements are required to bridge this gap.

\subsection{Ablation on Fine-tuning Datasets}

\begin{table}[htb]
 \begin{center}
 \resizebox{.75\width}{!}{
 \renewcommand{\arraystretch}{1.3}
 \begin{tabular}{l|llll}
 \hline\hline
 Model & B-U↑ & M-R↑ & R-L↑ & BERT-S↑ \\
 \hline
 & \multicolumn{4}{c}{MusicCaps}\\
\hline
MusiLingo / MusicCaps & \textbf{30.8} & 21.6& 21.7& \textbf{86.8}\\
MusiLingo / MI short & 2.1 & 8.4 & 9.0& 84.4\\
MusiLingo / MI long & 22.4 & \textbf{22.2} & \textbf{29.3}& 86.1\\
MusiLingo / MI mix & 20.4 & 20.2 & 27.2& 85.8\\
 \hline\hline
 & \multicolumn{4}{c}{MusicInstruct (Short)}\\
\hline
MusiLingo / MI(short) & \textbf{47.0}& \textbf{51.4}&\textbf{51.4}& \textbf{92.9}\\
MusiLingo / MI(long) & 7.2& 21.1& 56.5& 89.3\\ %505089
MusiLingo / MI(mixed) & 46.1& 50.9& 51.1& 92.8\\%505300
\hline\hline

 & \multicolumn{4}{c}{MusicInstruct (Long)}\\
\hline
MusiLingo / MI(short) & 12.3 & 13.6& 15.0& 83.2\\
MusiLingo / MI(long) & \textbf{45.0} & \textbf{25.0}& 22.9& \textbf{86.1}\\
MusiLingo / MI(mixed) & 40.3 & 24.3 & \textbf{23.6} & 85.6 \\\hline\hline

 & \multicolumn{4}{c}{MusicQA}\\ \hline
MusiLingo / MusicQA & \textbf{32.4} & \textbf{37.2} & \textbf{45.3} & \textbf{90.6}\\
MusiLingo / MI short & 27.6 & 34.0& 38.2 &  89.5\\
MusiLingo / MI long & 12.4 & 24.6 & \textbf{51.8} & 88.5\\
MusiLingo / MI mix  & 26.8  & 33.6  & 43.0  & 89.4 \\%505297
\hline\hline
 \end{tabular}
 }
\end{center}
 \caption{Ablition study results in MusiLingo performance after finetuning on a different partition of MI dataset.}
 \label{tab:ablition}
\end{table}
In this subsection, we present an ablation study that investigates the impact of fine-tuning datasets on the performance of MusiLingo, in the domain of music question answering. We explore how different fine-tuning strategies based on variations in training data, influence the effectiveness of MusiLingo. The fine-tuning datasets considered in our study are different partitions of MusicInstruct including MI (Short), MI (Long), and MI (all). 

Our investigation revealed that models trained on a combination of short objective questions and long subjective questions were consistently outperformed by models trained exclusively on a single partition of Q\&A pairs, even though we increased the calculation steps. This observation underscores the potential risk of incorporating diverse training data into the model training process, promoting enhanced performance. Besides, finetuning on MI (short) provides worse results on MI (long) and vice versa, suggesting a significant difference between short questions and long questions. Furthermore, we find that short questions are good for MusicQA zero-shot learning and long questions are good for captioning.

Overall, the results also highlight the importance of evaluating models in different scenarios to gain a more comprehensive understanding of their capabilities and limitations. This information can guide the development of more robust and versatile music question-answering systems in the future.

\section{Conclusion}
In summary, our submission introduces MusiLingo, a pioneering large language model that effectively bridges the gap between music and text domains. With the aid of a single projection layer, MusiLingo aligns music representations with textual contexts, delivering competitive performance in music captioning and question-answering tasks. The introduction of our innovative MusicInstruct dataset further enhances its capabilities. We envision that our work lays the foundation for a new era of multimodal applications in the field of music, offering exciting possibilities for both music enthusiasts and researchers, promising to revolutionise the way we engage with and comprehend music.

\section*{Limitations}
Our current model's fine-tuning process is relatively brief, and there is room for enhancing its performance through more extensive training and a more thorough exploration of hyperparameter configurations. Currently, the model provides good results on each dataset only after training on the same dataset and does not provide universality on all the downstream Q\&A datasets. We recognize these limitations and consider them as avenues for future research.

Furthermore, there might be some model hallucinations when GPT-4 generates the answer for long questions with subjective descriptions based on the input music, given the input to GPT only includes the annotation in the MusicCaps dataset and does not necessarily align with human feelings on the music excerpts. 

\section*{Ethics Statement}
Google has chosen to release only the YouTube IDs associated with the music in the MusicCaps dataset, refraining from providing the raw audio data. This approach introduces ambiguity regarding the dataset's copyright implications. Besides the audio, annotation is generated by AI algorithms -- the usage of GPT is to mimic human behaviour and we use it only for research use. We would like to emphasise that it cannot replace the human feeling towards music and we make our model public only for research use under a \texttt{cc-by-nc-sa} license. We acknowledge the need for transparent consideration of copyright ethics in dataset construction and use. %We require people only to use our dataset in a non-commercial way given the copyright issue.

\section*{Acknowledgement}
We would like to give thanks to Luca Marinelli and Ilaria Manco for the information on pre-training datasets. We would like to give thanks to Shansong Liu, one author of MU-LLaMA, for the suggestions and discussions. 

Yinghao Ma is a research student at the UKRI Centre for Doctoral Training in Artificial Intelligence and Music, supported by UK Research and Innovation [grant number EP/S022694/1]. 
Emmanouil Benetos is supported by a RAEng/Leverhulme Trust Research Fellowship [grant number LTRF2223-19-106].

The computations described in this research were performed using the Baskerville Tier 2 HPC service\footnote{\url{https://www.baskerville.ac.uk/}}. Baskerville was funded by the EPSRC and UKRI through the World Class Labs scheme (EP/T022221/1) and the Digital Research Infrastructure programme (EP/W032244/1) and is operated by Advanced Research Computing at the University of Birmingham.

This research was supported in part by the University of Pittsburgh Center for Research Computing, RRID: SCR\_022735, through the resources provided. Specifically, this work used the H2P cluster, which is supported by NSF award number OAC-2117681.

\bibliography{refs}

\newpage
\newpage

\appendix
\section*{Appendix}
\section{Prompt for MusicInstruct Q\&A Generation}
\begin{table*}[t]
\begin{tabular}{|l|l|}
\hline
\begin{tabular}[c]{@{}p{7.5cm}@{}}\textbf{v1:} (short Q\&A)\\ \textbf{System:}\\ You will be provided with a piece of caption that describes a music. The cation will be delimited with \#\#\#\# characters.\\ \\ Your task is to generate five question-answer pairs related to the music caption. The question should ask to describe the music content \textit{in detail}. The answer should be the answer to the question and contain details of the provided music caption.\\ \\ The question can include but not limited to any of the following information when the caption include them: music tempo, mood of the music, instruments used, singer, genre, music tags, or any inference, etc.\\ \\ IMPORTANT: Output a JSON object with the following four keys: 'Question 1', 'Answer 1', 'Question 2', 'Answer 2', 'Question 3', 'Answer 3', 'Question 4', 'Answer 4', 'Question 5', 'Answer 5'\\ \\ \textbf{User:}\\ \#\#\#\#\{query music\}\#\#\#\# \\ \\ \end{tabular} & \begin{tabular}[c]{@{}p{7.5cm}@{}}\textbf{v2:} (long Q\&A)\\ \textbf{System:}\\ You will be provided with a piece of caption that describes a music. The cation will be delimited with \#\#\#\# characters.\\ \\ Your task is to generate a conversational question-answer pair related to describing the music in detail. The question should ask to describe the music content \textit{in general}. The answer should be a paraphrased and well-structured paragraph based on the provided description, with a minimum of 100 words and a maximum of 200 words. The answer must be a paraphrased version of the provided information, very detailed and descriptive, and within the specified word count.\\ \\ \#\#SAMPLE QUESTIONS:\\ - Can you provide a summary of the music?\\ - What are the main features of the music?\\ - Could you briefly describe the music content?\\                         \\ IMPORTANT: Output a JSON object with only two keys: "Q" for question and "A" for answer.\\ \\ \textbf{User:}\\ \#\#\#\#\{query music\}\#\#\#\#\end{tabular}
\\ \hline
\end{tabular}
\caption{Example instruction used for prompting ChatGPT to generate Q\&A pairs based on a music caption.}
\label{eg}
\end{table*}

Example instructions used for prompting ChatGPT to generate the MusicInstruct dataset is given in Tab.~\ref{eg}.

\section{Example of MusicInstruct Q\&A}

The questions in MusicInstruct dataset include emotion, genre, singer, instrument information, tags, tempo etc. and human feelings towards the music clips. Such as typical music property that is interesting to music information retrieval. 

\subsection{version 1 - short Q\&A pairs}

~\\
Eg 1.

Q: What instruments are used in the song?

A: The song uses a piano, claps, kick hits, hi hats, and a synth bass. 

~\\
Eg 2.

Q: What genre does this music piece belong to?

A: This music piece belongs to the Japanese pop-punk genre.

~\\
Eg 3.

Q: What kind of media could this music piece be used in?

A: This music piece could be used in the soundtrack of a teenage/young adult feelgood movie/TV show or in the soundtrack of an animation movie or an anime.
~\\
\hrule
~\\
\subsection{version 2 - long Q\&A pairs}

~\\
Eg 1.

Q: Could you provide a detailed description of the church choir song?

A: The song is a prime example of a traditional church choir performance, one recorded in an era past. The arrangement sees the choir members joining their voices in an expressive harmony, imbued with a devout and contemplative essence. The pace of the song is moderated, neither too slow nor too fast, which enhances its immersive nature. Accompanying the choir is a church organ that amplifies the sacred ambiance without overshadowing the vocal chorus. Interestingly, there are no other musical instruments employed in the song. This speaks volumes about its pure and unadulterated nature. It's worth noting that the recording quality of the song is not up to par and it appears to have faded considerably over time, indicating it might have been a product of an amateur's handiwork at their residence. Despite preserving its classic church choir harmony essence, the deteriorated audio quality adds a nostalgic layer to the song, making it a noteworthy vintage piece in the world of devotional music. 

~\\
Eg 2.

Q: Could you describe the content and style of this song in detail?

A: The song in question is an amateur recording, featuring the unique art form of beatboxing. Beatboxing is a vocal technique primarily involving the artful mimicking of percussive instruments with one's mouth, voice, lips, tongue, and other parts. This specific rendition incorporates sounds representing the kick, snare, and hi-hat drum parts. Unlike typical songs, this one does not involve the use of any additional instruments or musical accompaniment. It presents an aural spectacle solely fueled by the beatbox performance. Moreover, there's an absence of a defining vocal melody in the conventional sense, as the core and the sole essence of the song lie in the nuanced and rhythmic tapestry of sounds created solely by beatboxing.

\section{Dataset Quality Evaluation Measure}\label{criteria}

Inspired by the evaluation in Kun~\cite{zheng2024kun}, we use the data quality evaluation criteria with 3 instruction evaluation aspects and 1 output evaluation described as follows. The instruction quality is assessed with a yes/unsure/no answer and the output quality is assessed with an excellent/pass/failed. 
The definition of criteria are:

\begin{enumerate}
    \item Instruction Clarity: Evaluators determine whether the instruction was unambiguous and coherent, encompassing necessary information without any vague terms or explanations. (y/u/n)
    \item  Instruction Feasibility: Evaluators assess whether the instruction was valid and answerable within the context and scope of the model’s capabilities. (y/u/n)
    \item Instruction Practicality: Evaluators judge the relevance of the instruction in music informatics scenarios. (y/u/n)
    \item Output Quality. The quality of the outputs is evaluated based on their alignment with the instructions. Evaluators are asked to rate each output as Excellent, Pass, or Fail, based on how well it met the requirements and intent of the instruction. (excellence/fair/fail.)
\end{enumerate}

All the annotators have instrument-playing experience but are not professional musicians. No ethical approval is needed for the dataset quality evaluation because we only invited authors for annotation.

\end{document}